\documentclass[9pt,twocolumn,twoside]{pnas-new}

\templatetype{pnasresearcharticle} 

\usepackage[normalem]{ulem}

\def\ka{{\bf k}}

\def\sx2{\sin{\frac{k_x}{2}}}
\def\sy2{\sin{\frac{k_y}{2}}}
\def\cx2{\cos{\frac{k_x}{2}}}
\def\cy2{\cos{\frac{k_y}{2}}}

\def\sqsx2{\sin^2{\frac{k_x}{2}}}
\def\sqsy2{\sin^2{\frac{k_y}{2}}}
\def\sqcx2{\cos^2{\frac{k_x}{2}}}
\def\sqcy2{\cos^2{\frac{k_y}{2}}}

\DeclareMathOperator\erfc{erfc}

\newcommand{\bns}{BaNiS$_2$}

\newcommand{\bcns}{BaCo$_{1-x}$Ni$_x$S$_2$} 
\newcommand{\michele}[1]{\textcolor{black}{\rm{#1}}}
\newcommand{\nilou}[1]{\textcolor{black}{\rm{#1}}}
\newcommand{\MF}[1]{\textcolor{black}{\rm{#1}}}
\newcommand{\AG}[1]{\textcolor{black}{\rm{#1}}}

\title{Moving Dirac nodes by chemical substitution}

\author[1]{Niloufar Nilforoushan}
\author[2]{Michele Casula}
\author[3,4]{Adriano Amaricci}
\author[1,5]{Marco Caputo}
\author[1]{Jonathan Caillaux}
\author[1,6]{Lama Khalil}
\author[1]{Evangelos Papalazarou}
\author[1]{Pascal Simon}
\author[7]{Luca Perfetti}
\author[3]{Ivana Vobornik}
\author[3,8]{Pranab Kumar Das}
\author[3]{Jun Fuji}
\author[3]{Alexei Barinov}
\author[2]{David Santos-Cottin}
\author[2]{Yannick Klein}
\author[4]{Michele Fabrizio}
\author[2]{Andrea Gauzzi}
\author[1]{Marino Marsi}

\affil[1]{Universit\'e Paris-Saclay, CNRS, Laboratoire de Physique des Solides,91405 Orsay, France.}
\affil[2]{Institut de Min\'eralogie, de Physique des Mat\'eriaux et de Cosmochimie (IMPMC), Sorbonne Universit\'e, CNRS UMR 7590, MNHN, 4 place Jussieu, 75252 Paris, France.}
\affil[3]{Istituto Officina dei Materiali (IOM) - CNR, Strada Statale
  14 km 163.5, 34149 Trieste, Italy.}
\affil[4]{International School for Advanced Studies SISSA, via Bonomea
  265, 34136 Trieste, Italy.}
\affil[5]{Sincrotrone Trieste, SS14 - Km 163.5, 34149 Trieste, Italy.}
\affil[6]{Synchrotron SOLEIL, Saint Aubin BP 48, Gif-sur-Yvette, F-91192, France}
\affil[7]{Laboratoire des Solides Irradi\'es, Ecole Polytechnique, CNRS, CEA, 91128 Palaiseau, France.}
\affil[8]{International Centre for Theoretical Physics, Strada Costiera 11, 34100 Trieste, Italy.}

\leadauthor{Nilforoushan} 

\significancestatement{
The on-demand control of topological properties with readily modifiable parameters is a fundamental step towards the design of novel electronic and spintronic devices. Here we show that this goal can be achieved in the correlated system \bcns\ , where we succeeded in significantly changing the reciprocal space position and shape of Dirac nodes by chemically substituting Ni with Co. We prove that the tunability of the Dirac states is realized by varying the electron correlation strength and the charge-transfer gap, both sensitive to the substitution level, $x$. Based on our finding, a class of late transition metal compounds can be established as prototypical for engineering highly tunable Dirac materials.
}

\authorcontributions{authors contribution: N.N., M.Cas., M.F., A.G. and M.M. designed research; N.N., M.Cas., A.A., M.Cap., J.C., L.K., E.P., P.S., L.P., I.V., P.K.D., J.F., A.B., D.S.C., Y.K., M.F., A.G. and M.M. performed research; N.N., M.Cas., A.A., M.F. and M.M. analyzed data; N.N., M.Cas., A.A.,M.F., A.G. and M.M. wrote the paper.}
\authordeclaration{author declaration: xxx.}

\correspondingauthor{To whom correspondence should be addressed. E-mail:niloufar.nilforoushan@phys.ens.fr ;   michele.casula@sorbonne-universite.fr ;  marino.marsi@universite-paris-saclay.fr}


\begin{abstract}
Dirac fermions play a central role in the study of topological phases, for they can generate a variety of exotic states, such as Weyl semimetals and topological insulators. The control and manipulation of Dirac fermions constitute a fundamental step towards the realization of novel concepts of electronic devices and quantum computation. By means of ARPES experiments and \emph{ab initio} simulations, here we show that Dirac states can be effectively tuned by doping a transition metal sulfide, BaNiS$_2$, through Co/Ni substitution. The symmetry and chemical characteristics of this material, combined with the modification of the charge transfer gap of BaCo$_{1-x}$Ni$_x$S$_2$ across its phase diagram, lead to the formation of Dirac lines whose position in $\ka$-space can be displaced along the $\Gamma - M$ symmetry direction, and their form reshaped.
Not only does the doping $x$ tailor the location and shape of the Dirac bands, but it also controls the metal-insulator transition in the same compound, making BaCo$_{1-x}$Ni$_x$S$_2$ a model system to functionalize Dirac materials by varying the strength of electron correlations.
\end{abstract}

\dates{This manuscript was compiled on \today}
\doi{\url{www.pnas.org/cgi/doi/10.1073/pnas.XXXXXXXXXX}}

\begin{document}

\maketitle
\thispagestyle{firststyle}
\ifthenelse{\boolean{shortarticle}}{\ifthenelse{\boolean{singlecolumn}}{\abscontentformatted}{\abscontent}}{}

\dropcap{I}n the vast domain of topological Dirac and Weyl materials \cite{Hasan2010a,bansil2016colloquium,Bradlyn2017,
Armitage2018,gibson2015three,vergniory2019high,Wang2012,Borisenko2014,liu2014stable}, the study of various underlying mechanisms \cite{muchler2012topological, Kane2005a, Young2012, Young2015, muechler2016topological, Yang2014} leading to the formation of non-trivial band structures is key to discover new 
topological electronic states \cite {Yan2017, Hubener2017, Bahramy2018, Fei2018, Cao2018, Cao2018a, Ye2018, Yin2018}. A highly desirable feature of these materials \AG{ is} the tunability of the topological properties by an external parameter, which will make them \AG{suitable in view of technological applications}, such as topological field effect transistors \cite{qian2014}. While a thorough control of band topology can be achieved in principle in optical lattices \cite{Tar12} and photonic crystals \cite{milicevic2019} through the wandering, merging and reshaping of nodal points and lines in $\ka$-space \cite{montambaux2018,gonccalves2019dirac}, in solid state systems such a control is much harder to achieve. Proposals have been made by using optical cavities \cite{hubener2020}, twisted van der Waals heterostructures \cite{kennes2020}, intercalation \cite{pakhira2018}, chemical deposition \cite{tsujikawa2020,roy2014}, impurities \cite{miao2018}, and magnetic and electric applied fields \cite{galanakis2012}, both static \cite{diaz2017} and time-periodic \cite{diaz2019,Hubener2017}. Here, we prove that it is possible to move and reshape Dirac nodal lines in reciprocal space \michele{by chemical substitution}. 
\AG{Namely,} by means of Angle Resolved Photo-Emission Spectroscopy (ARPES) experiments and \emph{ab initio} simulations, we observe a sizable shift of robust massive Dirac nodes towards $\Gamma$ in \bcns\, as a function of doping $x$, obtained \michele{by replacing Ni with Co}. At variance with previous attempts of controlling Dirac states by doping \cite{zhou2012,Fei2018}, in our work we report both a reshape and a significant $\ka$-displacement of the Dirac nodes.

\begin{figure*}[hbt]
\centering
\includegraphics[width=17.5cm]{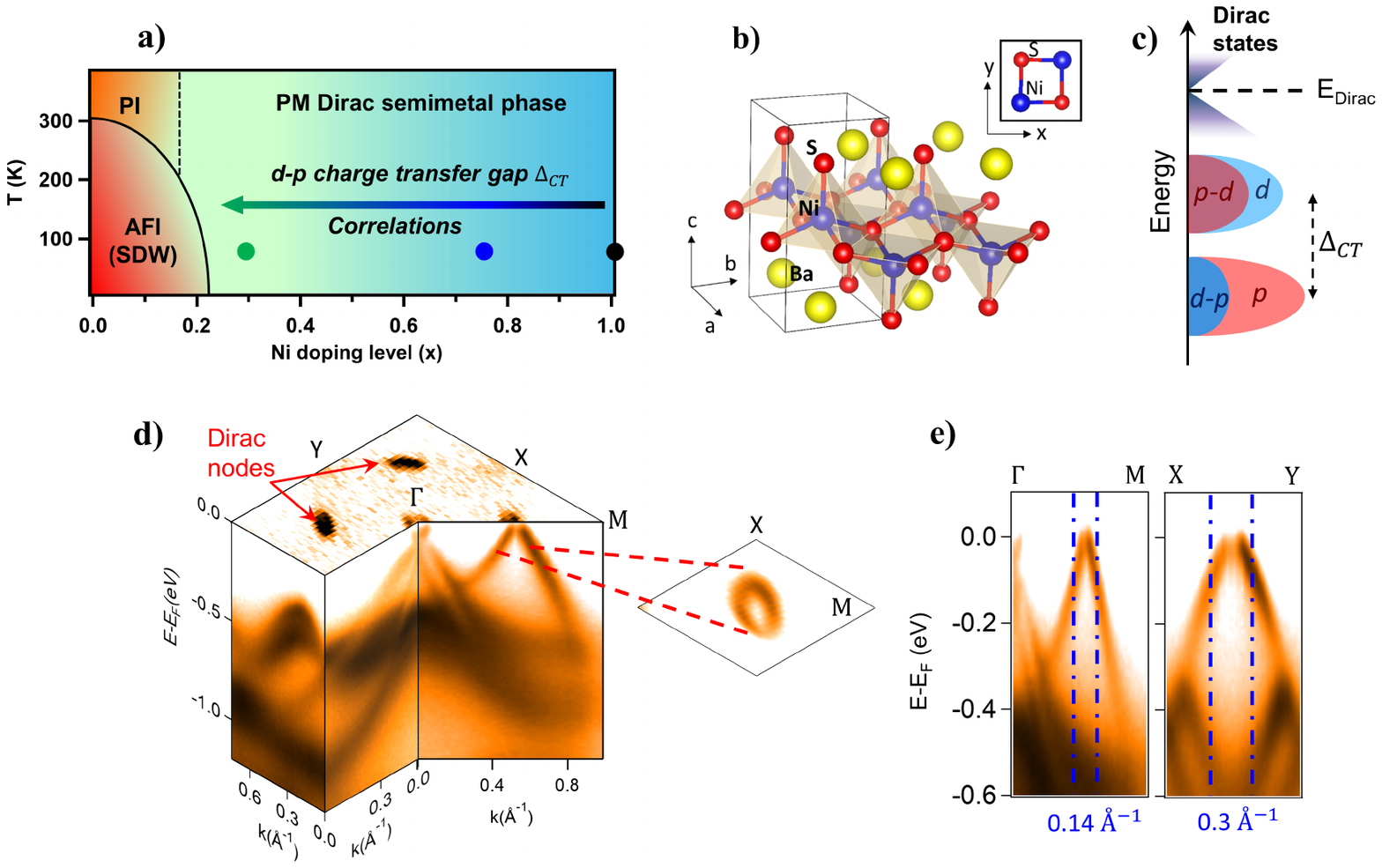}
\caption{Experimental observation of Dirac states in the phase diagram of BaCo$_{1-x}$Ni$_x$S$_2$.
  (a) Phase diagram of BaCo$_{1-x}$Ni$_x$S$_2$. The transition lines between the paramagnetic metal (PM), the
  paramagnetic insulator (PI), and the antiferromagnetic insulator
  (AFI) are reported.
Colored circles
  indicate the different doping levels $x$ studied in this work. This doping alters
  the $d-p$ charge transfer gap ($\Delta_{CT}$). 
  (b) Crystal structure of BaNiS$_2$. Blue, red and yellow spheres
  represent the Ni, S and Ba atoms, respectively. The tetragonal unit
  cell is indicated by black solid lines. Lattice parameters are $a$ =
  4.44 \AA~ and $c$ = 8.93 \AA \cite{Grey1970}.
  Top: projection of the unit cell in the $xy$ plane, containing two Ni
  atoms.
 (c) Schematics of the energy levels. The hybridization of $d-$ and $p-$ orbitals creates the Dirac states, and the $d-p$ charge transfer gap fixes the position of these states in the $E-k$ space. (d) A three-dimensional ARPES map
  of BaNiS$_2$ ($x=1$) taken at
  70 eV photon energy. The top surface shows the Fermi surface and the
  sides of the cube present the band dispersion along high-symmetry
  directions . The linearly dispersing  bands along 
  $\Gamma - M$ cross each other at the Fermi level, $E_F$, thus
  creating four Dirac nodes.
  (e) We observe the
  oval-shaped section of the linearly dispersing bands on the
  $k_x-k_y$ plane for $E-E_F= -100$ meV.
  The linearly dispersing bands along the 
  major and minor axis of the 
  oval are also shown.}
\label{fig:fig1}
\end{figure*}

\bcns\ is a 
prototypical
transition metal system with a simple square lattice \cite{Sato2001}.
In \bcns\, the same doping parameter $x$ that tunes the position of the Dirac nodes also controls the \AG{electronic} phase diagram, 
which features a \michele{first-order} metal-insulator transition (MIT) at
a critical substitution level, $x_{cr}\sim$ 0.22 \cite{Martinson1993a,Krishnakumar2001}, as shown in
Fig.~\ref{fig:fig1}(a). The Co-rich side ($x=0$) is \michele{an insulator with collinear magnetic order and with local moments in a high-spin (S=3/2) configuration \cite{Mandrus1997}}. Both electron
correlation strength and charge-transfer gap $\Delta_{CT}$ increase with decreasing $x$, as typically found in the late transition metal series. The MIT at $x=0.22$ is of interest because it is 
driven by electron correlations \cite{SantosCottin2018} and is associated with a competition
between an insulating antiferromagnetic phase and an unconventional
paramagnetic semi-metal \cite{SantosCottin2016b}, where the Dirac nodes are found at the Fermi level.
We show that
a distinctive feature of these Dirac states
is
their
dominant 
$d$-orbital character \AG{and} that the underlying band 
inversion mechanism is driven by a large $d-p$ hybridization combined
with the non-symmorphic symmetry (NSS) of the crystal (see
Fig.~\ref{fig:fig1}(b)). It follows that an essential role in controlling the properties of Dirac states is played by electron correlations and by the charge-transfer gap (Fig. \ref{fig:fig1} (c)), as they have a direct impact on the hybridization strength. 
This results into an effective 
tunability of shape, energy and wave vector of the Dirac lines in the proximity of the Fermi level. \AG{Specifically, the present ARPES study unveils Dirac bands moving from $M$ to $\Gamma$ with decreasing $x$.} The bands are well explained quantitatively by
\textit{ab initio} calculations, in a hybrid density functional approximation suitable for including \michele{non-local \MF{correlations} of screened-exchange type, which affect the hybridization between the $d$ and $p$ states. The same functional is able to describe the insulating spin-density wave (SDW) phase at $x=0$, driven by local correlations, upon increase of the optimal screened-exchange fraction.}
These calculations confirm
that the Dirac nodes mobility in $\ka$-space 
stems directly from the evolution of the charge transfer gap, i.e. the relative position between
$d$ and $p$ on-site energies. These results clearly suggest that
BaCo$_{1-x}$Ni$_x$S$_2$ is a model system to tailor Dirac states and,
more generally, that two archetypal features of correlated systems
such as the hybrid $d-p$ bands and the charge-transfer gap constitute 
a promising playground to engineer Dirac and topological materials
using chemical substitution and other macroscopic control parameters.

\section*{Observation of Dirac states in BaNiS$_2$}
\label{electronic_structure}
We begin with the undoped sample BaNiS$_2$. In Fig.~\ref{fig:fig1}(d),
we represent a three dimensional ARPES map of the Brillouin zone (BZ) for the
high symmetry directions. Along $\Gamma - M$, we observe linearly
dispersing bands and -within ARPES resolution- gapless nodes at the Fermi level $E_F$. 
The Fermi surface reveals two pairs of such Dirac-like crossings related to each
other by the time-reversal and by the two-fold rotation axis $C_2$ of
the $C_{2v}$ little group for the $\ka$-vectors along 
$\Gamma - M$. The Dirac nodes lie on the $\sigma_d$ reflection planes
and extend along the $k_z$ direction piercing the whole BZ, unlike other topological node-line
semimetals known to date, like Cu$_3$NPd \cite{Kim2015,yu2015topological}, Ca$_3$P$_2$ \cite{Xie2015} and ZrSiS \cite{Schoop2016}, where the nodal 
lines
form closed areas around high-symmetry points.

As one can see in Fig.~\ref{fig:fig1}(e), along the $\Gamma - M$
direction, the linearly dispersing bands remain isolated up to about
$E-E_F = -0.35$ eV. 
These bands create an oval-shaped
section on the constant energy maps near the Fermi level (see 
Fig.~\ref{fig:fig1}(d)). This asymmetry, clearly visible in
Fig.~\ref{fig:fig1}(e), arises from the tilted type-I nature of the
Dirac cone. 
The model Hamiltonian explaining the low-energy
spectrum of the linearly dispersing bands observed experimentally is described in the Supporting Information (SI), c.f. Sec.~S1.
The linear bands present no \AG{$k_z$ 
dispersion,} as shown in the SI (Sec.~S2 and Fig.~S2). The absence of dispersion is an indication of the 2D nature of the Dirac cones. The Dirac point remains pinned almost at the Fermi level - about 30 meV above - and its wave vector is fixed along the $\Gamma (Z) - M (A)$ direction. Here, the Dirac point position is obtained by extrapolating the ARPES data. These values are in perfect agreement with those directly measured in a recent pump-probe experiment \cite{Nil2020}.

\begin{figure*}[htb]
\centering 
\includegraphics[scale=0.5]{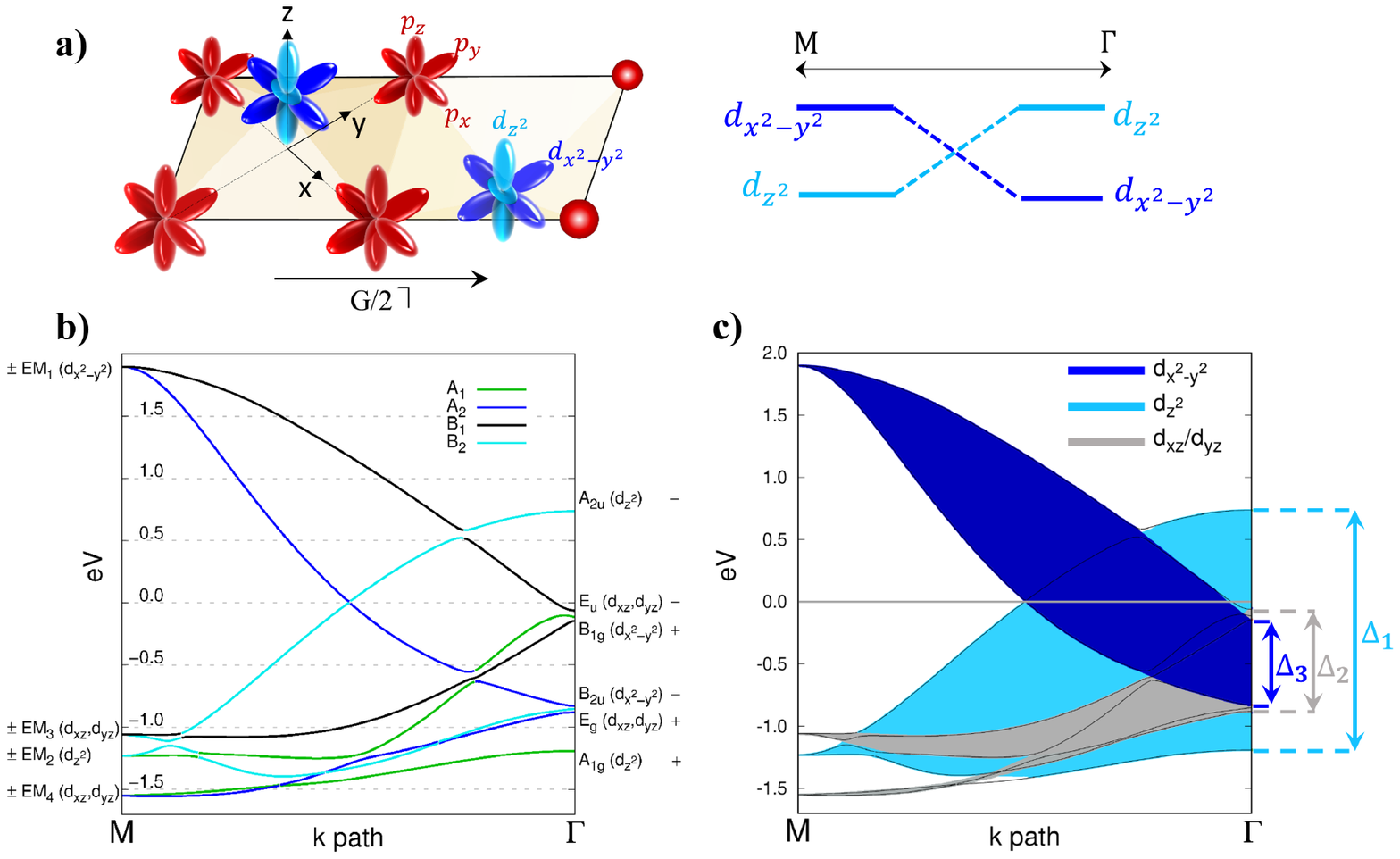}
\caption{Mechanism of band inversion and formation of
    hybridized Dirac states. (a) Schematics of the $d$- and
  $p$-orbitals of Ni and S, respectively. Right panel: The strong
  hybridization of the $d$-orbitals with the ligand $p$-orbitals,
  favored by the NSS, is responsible for the band inversion. (b) Band symmetries along the M$\Gamma$ direction. At the 
right (left) -hand side of the panel, we report the symmetries at 
$\Gamma$ (M), while the symmetries in between follow the \michele{irreducible representations} of the 
$C_{2v}$ point group, represented by the color code in the 
key. The outer $+/-$ signs indicate the parity of the respective Bloch 
wave functions at the beginning and at the end of the $\ka$-path. 
(c) Evolution of the energy splitting between even and 
odd combinations of $d$-orbitals along M$\Gamma$. The dominant orbital character is reported.
The navy blue (blue) vertical arrows indicate the 
splitting $\Delta_3$ ($\Delta_1$) between the $d_{x^2-y^2}$ ($d_{z^2}$) bands at $\Gamma$ due 
to the hybridization with the ligand $p_z$-orbitals.  
The grey arrow 
indicates the splitting $\Delta_2$ of $d_{xz}$/$d_{yz}$ bands at $\Gamma$ due to their hybridization 
with the $p_x$/$p_y$ orbitals.}
\label{fig:Figure_C2v}
\end{figure*}

\section*{Symmetry analysis of the electronic bands: mechanism of band inversion and formation of Dirac states}
\label{symmetries}

To unveil the physical mechanism responsible for the formation of
Dirac cones in \bns\ we performed a detailed theoretical analysis of
the symmetry of the electronic bands. We carried out density functional theory (DFT) calculations, by employing
  \michele{a modified} Heyd–Scuseria–Ernzerhof (HSE) functional. The details of the band structure are presented in Methods and SI (see SI Sec.~S3, where 
  \michele{we also discuss how the inclusion of the spin-orbit coupling (SOC) affects the topological properties}).
  The use of the HSE functional is dictated by 
  non-local 
  correlation effects present in this material. Indeed,
  a hybrid HSE functional 
  \michele{with the optimal screened-exchange fraction $\alpha=7\%$ (see Eq.~\ref{HSE}) is} 
  needed to
  account for the Fermi surface renormalization of BaNiS$_2$ seen in
  quantum oscillations \cite{Klein2018}. Previous theoretical calculations
\cite{Hase1995,Krishnakumar2001,SantosCottin2016} have shown that both
S 3$p$- and Ni 3$d$-orbitals contribute to the Bloch functions near
the Fermi level. We ascribe the electronic states close to the Fermi level mainly to
the Ni $3d$-orbitals hybridized with the S $3p$-orbitals. 
\michele{In this situation, the exchange contribution to the hybridization with the ligands plays a crucial role in determining the topology of the Fermi surface (Fig. S6(b) illustrates the electronic structure dependence upon $\alpha$).}
Hereafter,
we consider a Cartesian reference frame where the $x$- and $y$-axis
are parallel to the Ni-S bonds in the tetragonal
$ab$-plane. Neighbouring Ni ions are aligned along the diagonal $xy$
direction (Fig.~\ref{fig:fig1}(b)). In this frame, at the crossing
points, located along the $(u,u,v)$ directions, the bands have
dominant $d_{z^2}$ and $d_{x^2-y^2}$ character. This multi-orbital
nature was confirmed by a polarization dependent laser-ARPES study
(see SI, Sec.~S4).

As sketched in Fig.~\ref{fig:Figure_C2v}(a), the crystal structure of
\bns\ is made of square-lattice layers of staggered, edge-sharing
NiS$_5$ pyramids pointing along the out-of-plane [001] $c$-axis
direction \cite{Martinson1993a}. The Ni atoms inside the S pyramids
probe a crystal field that splits the atomic $d$-shell into the
following levels (in descending energy order): $d_{x^2-y^2}$,
$d_{z^2}$, the degenerate doublet $(d_{xz},d_{yz})$ and $d_{xy}$. Due
to the $3d^8 4s^0$ electronic configuration of the Ni$^{2+}$ ion, we
expect all $d$-orbitals to be filled, except the two highest ones,
$d_{x^2-y^2}$ and $d_{z^2}$, which 
are nearly half-filled assuming that the Hund's exchange is sufficiently strong.

The puckering of the BaNiS$_2$ layers gives rise to a tetragonal nonsymmorphic $P$4/$nmm$ structure characterized by a horizontal gliding plane which generates two Ni and two apical S positions at (1/4,1/4,$z$) and (3/4,3/4,$-z$), separated by a fractional {\bf f}=(1/2,1/2,0) translation in the plane, Fig.~\ref{fig:Figure_C2v}(a). The two Ni atoms occupy Wyckoff position $2c$, corresponding to the \textit{M} \AG{ symmetry}, while the two planar S are at the $2a$ site  corresponding to the $\Gamma$ \AG{ symmetry}.

At \textit{M}, the energy hierarchy of the atomic orbitals follows
closely the crystal field splitting
(Fig.~\ref{fig:Figure_C2v}(b)). The little group admits the following
four 2D irreducible representations \nilou{(irreps)} $E_{Mi=1,\dots,4}$ \cite{Vafek-PRB2013}, each originating from the same orbitals of the
two inequivalent Ni.
However, the levels stacking at $\Gamma$, whose little group is isomorphic to $D_{4h}$, differs from that
predicted by the crystal field. This is due to the sizable hybridization of Ni $d$-orbitals with the S $p$ ligands (see SI, Sec.~S5). Owing to the NSS, each Bloch eigenfunction at $\Gamma$ is either even or odd upon exchanging the inequivalent Ni and S within each unit cell. Even and odd combinations of identical $d$-orbitals belonging to inequivalent Ni atoms split in energy since they hybridize differently with the ligands. 
The even combination of the $d_{x^2-y^2}$ Ni orbitals is weakly
hybridized with the $p_z$-orbitals of the planar S, since the two Ni atoms
are out of the basal plane. On the other hand, the odd combination is
non-bonding. It follows that the $B_{1g}$ even combination shifts up
in energy with respect to the $B_{2u}$ odd one. Seemingly, the
$A_{2u}$ odd combination of the $d_{z^2}$-orbitals hybridizes
substantially with the $p_z$-orbitals of the planar and apical S, thus
increasing  significantly the energy of the odd
combination. Eventually, its energy raises above
 the $B_{1g}$ and $B_{2u}$ levels, as well as 
the $A_{1g}$ state (even combination of $d_{z^2}$-orbitals).
This leads to a reverse of the crystal field order as reported in Fig.~\ref{fig:Figure_C2v}(c). 
 
Because the irreps at the \textit{A} and \textit{Z} $\ka$-points are equivalent to those at \textit{M} and $\Gamma$ \cite{Vafek-PRB2013}, respectively, the orbital hierarchy found at \textit{M} and $\Gamma$ must be preserved along the $M - A$ and $\Gamma - Z$ directions. Thus, for any $v$ along the $(0,0,v) \rightarrow (1/2,1/2,v)$ path, 
a band inversion between bands with predominant $d_{z^2}$ and
$d_{x^2-y^2}$ characters must occur. Therefore, band crossing is allowed without SOC, and leads to two
Dirac points at a given $k_z$ right at the Fermi energy for $k_z=0$. Indeed, the
crossing bands transform like different irreps of the little group,
which is isomorphic to $C_{2v}$ for a $\ka$-point $(u,u,v)$ with
$v=0,1/2$, and to $C_s$ with $v\in ~ ]0,1/2[$. 
These Dirac nodes are massive as a consequence of the SOC, which makes the material a weak topological insulator. The SOC gap is however very small (about 18 meV), 
\nilou{and} below
ARPES resolution. Nevertheless, the focus of the present work is not on these very-low-energy features, but rather on the tunability of the whole Dirac nodal structure. In the family of weak topological insulators having the same $P4/nmm$ space group and showing SOC gapped Dirac cones along the $\Gamma-M$ direction (such as ZrSiS, for instance), 
\bcns\ is a peculiar member. Indeed, the strong local Hund's exchange coupling favors nearly half-filled $d_{x^2-y^2}$ and $d_{z^2}$ orbitals, that explains the proximity of the Dirac nodes to the Fermi level for $x=1$,
in accordance also to Luttinger's theorem (see SI, Sec.~S6).
This is another \michele{signature} of the relevance of 
electron correlations in this transition metal compound, \michele{which manifest themselves in both local and non-local contributions, the former leading eventually to the insulating phase at the Co side of BaCo$_{1-x}$Ni$_x$S$_2$, the latter affecting the variation of $\Delta_{CT}$ across the series.}


\section*{ARPES evidence of Dirac states tuned by doping, $x$}
\begin{figure*}
\centering
\includegraphics[scale=0.75]{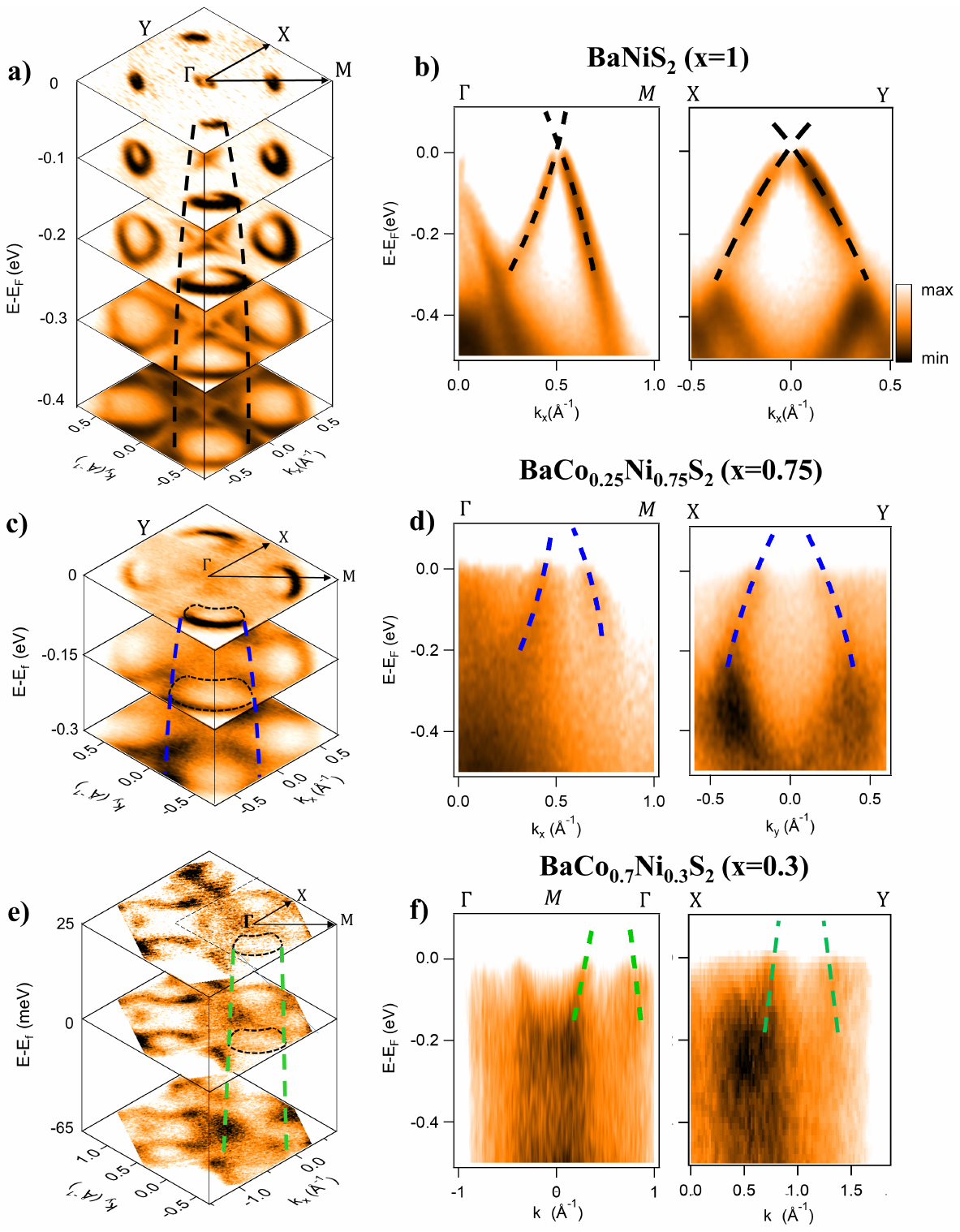}
\caption{Experimental ARPES evolution of Dirac states with
    doping $x$, in \bcns. (a, b) Dispersion of these states
    for $x=1$. Panel (a) shows the iso-energy contours with increasing
    binding energy. Panel (b) shows the dispersion along the high
    symmetry directions $\Gamma - M$ and $X - X^{\prime}$. Note the
    anisotropy of the dispersion, which is due to the oval shape of
    the pockets at the Fermi surface. Dashed lines are a guide to the
    eye that represent schematically the dispersion. (c-d) and (e-f):
    the same as in panels (a-b) for the $x=0.75$ and the $x=0.3$
    samples, respectively. All spectra are obtained with a photon
    energy of 70 eV.}
\label{fig:fig3}
\end{figure*}

\begin{figure*}[htb]
\centering
\includegraphics[width=\textwidth]{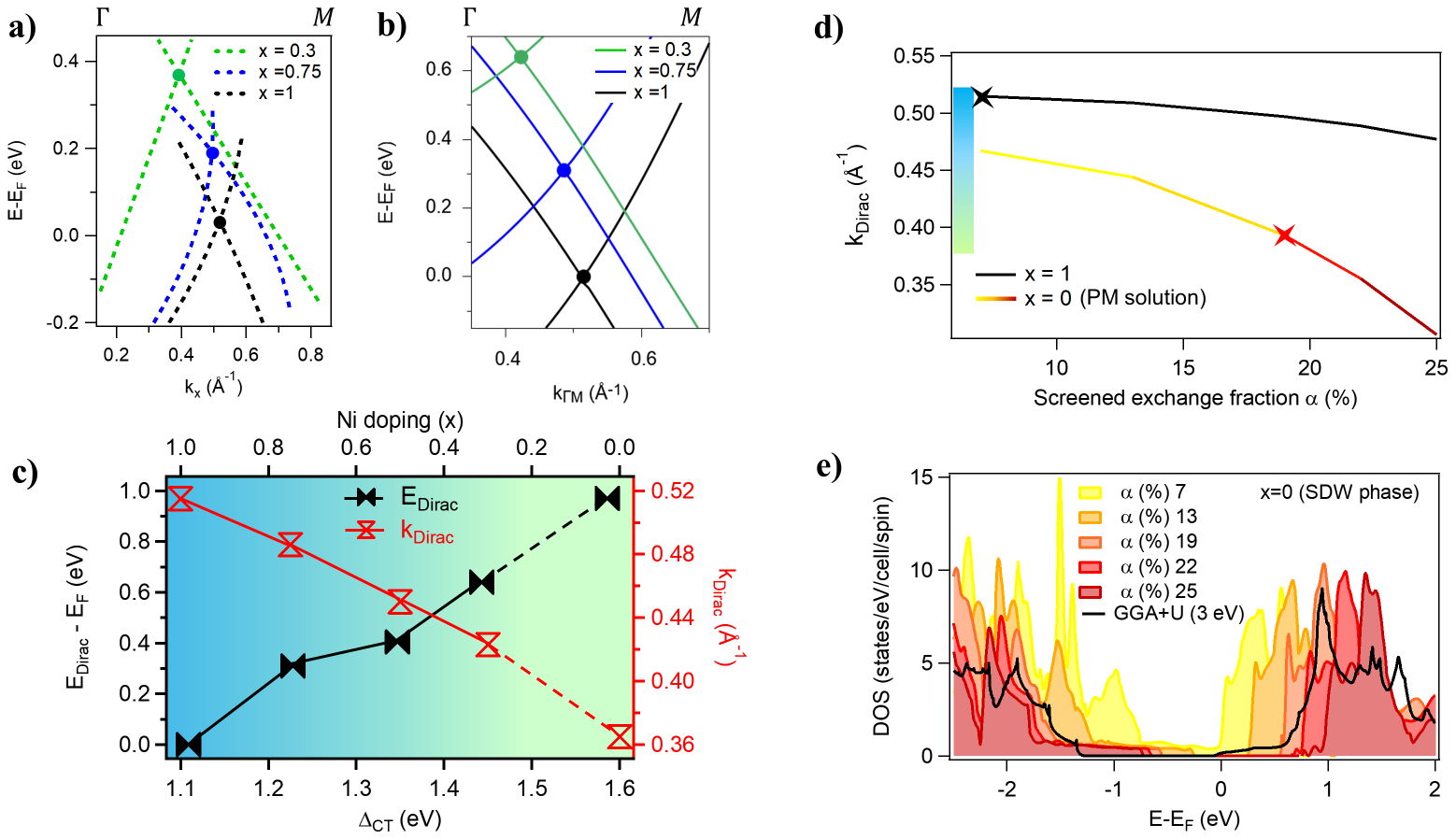}
\caption{Evolution of Dirac states with doping x, in
    BaCo$_{1-x}$Ni$_x$S$_2$. (a) 
Curves fitting the experimental Dirac states for different values of $x$. (b) 
Evolution of the Dirac states predicted by our HSE/tight-binding
calculations of the band structure. (c) Variation of the charge transfer gap, $\Delta_\textrm{CT}$, with
the Ni content, $x$. The point at $x=0$ is the metastable state adiabatically connected with the metallic phase found for $x>x_{cr} \simeq 0.22$. By reducing $x$ from 1 to 0, the Dirac point,
$\ka_\textrm{Dirac}$, moves towards $\Gamma$. Correspondingly, the
difference between the energy of this point and the Fermi level
increases. (d) $\ka_\textrm{Dirac}$ variation for $x=0$ and $x=1$, when we relax the screened-exchange fraction parameter $\alpha$ in the modified HSE functional. The black (red) star is located at the value of $\alpha$ that optimally captures the correlation strength in BaNiS$_2$ (BaCoS$_2$), used to predict the evolution in panels (b) and (c). (e) Density of states (DOS) of the $x=0$ collinear SDW solution, as computed by GGA+U (black line) with \emph{ab initio} values for the local Hubbard repulsion, and the modified HSE with values for $\alpha$ covering different correlation strengths. Optimal $\alpha=19\%$ best matches the GGA+U DOS.
}
\label{fig:fig4}
\end{figure*}

We now turn our attention to the effect of the Co/Ni substitution on the
evolution of the band structure, notably the Dirac states. 
According to the \bcns\ phase diagram, this substitution modifies the strength of the electron-electron correlations and the amplitude of $\Delta_{CT}$.
A series of ARPES
spectra are given for the $x=0.75$ and  $x=0.3$ compositions. In
Fig.~\ref{fig:fig3}, we display the evolution of the Fermi surface and
the electronic band structure along $\Gamma-M$ with $x$. For $x=0.75$,
the Fermi surface is composed of a four-leaf feature at the $\Gamma$
point and four hole-like pockets along the $\Gamma-M$,
Fig.~\ref{fig:fig3}(c). These pockets originate from the Dirac states
crossing the Fermi level. The Dirac cone is shown in
Fig.~\ref{fig:fig3}(d) along and perpendicular to the $\Gamma - M$
direction. At higher substitution levels, for $x=0.30$, the Dirac
states shift  up to lower binding energies,
so the size of the hole-like pockets in the $k_x-k_y$ plane is increased
(see Fig.~\ref{fig:fig3}(e,f)). The ARPES signal is also broader: since 
our structural study indicates that
the crystalline quality is not affected by Co/Ni substitution (see Sec.~S7 and Tab.~S2 in SI), 
this broadening is consistent with the increase in electron-electron correlations
while approaching the metal-insulator transition. 
\cite{Hase1995,Sato2001,SantosCottin2018}. On the theoretical ground, this is expected because \michele{Co-substitution} brings the whole $d$-manifold closer to 
  fillings where \michele{local} correlation effects are enhanced,
  according to the Hund's metals picture \cite{deMedici2011}.
Fig.~\ref{fig:fig4}(a) schematically illustrates the evolution of the Dirac cone with $x$; in
Table \ref{table:Dirac_point} we give the position of the Dirac points
determined by extrapolating the band dispersion. In summary, one notes
that the Co-substitution \michele{moves the Dirac points further beyond the Fermi level} and reduces its wave vector.

\begin{table}[h]
\centering
\caption{Position of the Dirac point (DP) in energy and momentum space for different doping levels in the metallic phase.}
\label{table:Dirac_point}
\begin{center}
\begin{tabular}{c c c}
\hline
\hline
\makebox[60pt][c]{Compound} & \makebox[40pt][c]{E$_{DP}$ (eV)} & \makebox[40pt][c]{$k_{DP}$(\AA$^{-1}$)} \\
\hline
\makebox[60pt][c]{BaNiS$_{2}$} & 0.03$\pm$ 0.01 & 0.52$\pm$ 0.01\\
\makebox[60pt][c]{BaCo$_{0.25}$Ni$_{0.75}$S$_2$} & 0.19$\pm$ 0.01 & 0.49$\pm$ 0.01\\
\makebox[60pt][c]{BaCo$_{0.7}$Ni$_{0.3}$S$_2$} & 0.37$\pm$ 0.02 & 0.39$\pm$ 0.02\\
\hline
\hline
\end{tabular}
\end{center}
\end{table}

\section*{Evolution of Dirac states with doping}

In order to account for the tunability of the Dirac cones detected by ARPES, 
\michele{we carried out extensive \textit{ab initio} DFT-HSE calculations as a function of the screened-exchange fraction $\alpha$, which controls the correlation strength in the modified hybrid functional framework. To explicitly include the charge transfer $\Delta_{CT}$ variation led by chemical substitution, we computed the two end-members of the BaCo$_{1-x}$Ni$_x$S$_2$ series, namely $x=1$ (BaNiS$_2$) and $x=0$ (BaCoS$_2$). For \MF{$x=1$} the optimal $\alpha=7\%$, since it reproduces the frequencies of quantum oscillations in BaNiS$_2$ \cite{Klein2018}. In order to fix such percentage for \MF{$x=0$}, we performed \emph{ab initio} calculations assuming the collinear SDW observed experimentally \cite{Mandrus1997}, by means of both HSE and the generalized gradient approximation supplemented by local Hubbard interactions (GGA+U). The strength of the Hubbard repulsion $U=3.0~\text{eV}$ and local Hund's coupling $J=0.9~\text{eV}$, included in GGA+U, was estimated from first principles within the constrained random phase approximation \cite{SantosCottin2018}. GGA+U correctly predicts an insulating state (Fig.~\ref{fig:fig4}(e)). By varying the percentage $\alpha$ of screened exchange in HSE, we find that, while $\alpha=7\%$ gives a metal, $\alpha\simeq 19\%$ reproduces the main peaks across gap obtained by GGA+U (Fig.~\ref{fig:fig4}(e)). This result suggests that HSE can describe BaCo$_{1-x}$Ni$_x$S$_2$ only if the percentage of screened exchange $\alpha$ increases from 7\% up to around 19\% with decreasing $x$ from 1 to 0. Starting from the most correlated Co side, the reduction of the Hubbard repulsion upon electron doping, implied by the $\alpha$ dependence on $x$, has been found in other strongly correlated compounds, such as La-doped Sr$_2$IrO$_4$ \cite{Liu2016}.}

\michele{In BaCoS$_2$, beside the SDW solution compatible with the \MF{observed low-temperature state}, it is possible to obtain another one \MF{once} magnetism is not allowed, 
namely \MF{forcing spin $SU(2)$ symmetry}. This paramagnetic metallic (PM) phase is metastable at low temperature, and adiabatically connected with the metallic solution at $x=1$. Therefore, it hosts Dirac cones; it is metallic and separated by \MF{an energy barrier} from the stable insulating SDW phase.} \michele{In Fig.~\ref{fig:fig4}(d), we plot the distance of the Dirac node ($\ka_\textrm{Dirac}$) from the $\Gamma$ point as a function of $\alpha$, for $x=1$ and the metallic solution at $x=0$. 
$\ka_\textrm{Dirac}$ strongly depends on both $x$ and $\alpha$ (See sec.~S8, Figs.~S6(a) and S6(b) plot the band structures where the $\ka_\textrm{Dirac}$ values have been extracted from). 
By taking the optimal $\alpha$'s for each $x$, the Dirac node is predicted to drift from $\ka_\textrm{Dirac} \simeq 0.52\,\text{\AA}^{-1}$ at $x=1$ down to $\ka_\textrm{Dirac}\simeq 0.38\,\text{\AA}^{-1}$ at $x=0$, covering the colored $y$-axis range in Fig.~\ref{fig:fig4}(d), in agreement with the range of variation seen in experiment.}

\michele{Next, we analyze the 22-bands full $d-p$ tight-binding model derived from the \textit{ab initio} DFT-HSE for $x=1$ (with $\alpha=7\%$) and for $x=0$ (with $\alpha=19\%$), c.f. Sec. S9, and Fig. S8} The $x=0$ state has shifted Dirac cones in both $\ka$ and energy position with respect to the BaNiS$_2$ parent compound. To underpin the mechanism
behind the evolution of the cones, 
we compared the two tight-binding Hamiltonians for $x=0$ and $x=1$
The main difference involves the on-site energies and, in particular,
the relative position of the $p$ and $d$ states, i.e. the charge
transfer gap $\Delta_{CT}$.  
This proves that the doping $x$ \michele{via chemical substitution} is indeed an effective control parameter, as it alters the $d - p$ charge transfer gap $\Delta_{CT}$ \michele{together with the correlation strength} and, consequently, 
the $d - p$ hybridization amplitude, 
which directly affects
position and shape of the Dirac nodes.

\michele{In the following, we define}
$\Delta_{CT}$ as the energy difference between \michele{the average energy position of the full $d$ manifold and the average one of the $p$ manifold}.   
According to our HSE
calculations, $\Delta_{CT}$ varies from 1.1 eV ($x=1$) to 1.6 eV
($x=0$). Assuming a linear variation of $\Delta_{CT}$ and on-site
energies upon \michele{Ni-content $x$}, we are able to estimate
$\Delta_{CT}=\Delta_{CT}(x)$ and, thus, predict the evolution of the
band structure and Dirac states \michele{by interpolating between the BaCoS$_2$ and BaNiS$_2$ TB models}. This evolution is reported in Fig.~\ref{fig:fig4}(b), while
the actual Dirac states dynamics - represented by the behavior of both
the $\ka$ and energy position of the Dirac point as a function of
$\Delta_{CT}$ - is plotted in Fig.~\ref{fig:fig4}(c). 
This 
shows
that the tunability upon doping found 
experimentally does not merely consist of a rigid shift
of the Dirac cones \cite{Fei2018}, but it
involves the change of both their shape and $\ka$-position (see also Fig.~S8).

This theoretical prediction is in good agreement with the observed evolution of the Dirac
cone with $x$,
as apparent in Fig.~\ref{fig:fig4}(a). Such movable Dirac nodes in the $\ka$-space
have recently attracted a great deal of interest from theory \cite{Yang2014,gonccalves2019dirac, Car13}, as well as in the context of optical lattices \cite{Tar12} and photonic crystals \cite{milicevic2019}. 
The present system offers the opportunity of observing in a real material how a simple experimental parameter - chemical \michele{substitution} - can be used to tune Dirac states. 

Manipulating the shape and position of the Dirac cones is also expected in \bcns\ using pressure in bulk samples or strain in thin films. Specifically, strain can be used to distort the 
square lattice, thus breaking one of the symmetries that protect the fourfold 
Dirac nodal lines. Non-trivial phases, such as Weyl semimetals, could then be triggered by time-inversion breaking perturbations, like an external electromagnetic field. A further possibility is the creation of spin-chiral edge states thanks to the proximity of the material to a topological insulator.

\section*{Conclusion}
In conclusion, \AG{we have shown that \bcns\ offers the opportunity of effectively tuning Dirac bands by exploiting a peculiar inversion mechanism of $d$-electron bands.} \AG{Namely, the Co/Ni substitution has been found to alter both the charge transfer gap and 
the strength of the electron-electron correlations that control position and shape of the bands}. Remarkably, the same Co/Ni substitution makes it possible to span the \AG{electronic} phase diagram, with the Dirac states present across its metallic phase. 
We emphasize the \AG{applicability of the present approach} to a wide class of materials described by the $d - p$
effective Hamiltonian, thus enabling to forge new Dirac states controlled by \michele{chemical substitutions. This opens the perspective of} \AG{engineering} Dirac states in correlated \AG{electronic systems} \michele{by exploiting macroscopically tunable parameters}.

\matmethods{
\subsection*{ARPES measurements} Single crystals of
BaCo$_{1-x}$Ni$_x$S$_2$ were cleaved \textit{in-situ}, exposing the \textit{ab} plane under UHV conditions (base pressure better than
$10^{-11}$ mbar). Most of the synchrotron radiation ARPES measurements were performed on the
Advanced Photoelectric Effect (APE) beamline at the Elettra light source, 
with linearly polarized beam and different photon energies. The sample
temperature was 70 K. The data were collected with a VG-DA30 Scienta hemispherical analyzer that
operates in deflection mode and provides high-resolution
two-dimensional ${\bf k}$-space mapping while the sample geometry is fixed \cite{Bigi2017}. The total measured energy resolution is $\sim$ 15 meV and the angular resolution is better than 0.2$^{\circ}$.
Some of the data were also acquired with a 6.2 eV laser source \cite{Caputo2018}; and some at the Spectromicroscopy beamline \cite{Dudin2010}: the end station hosts two exchangeable multilayer-coated Schwarzschild objectives (SO) designed to focus the radiation at 27 eV and 74 eV to a small spot ($\sim$600 nm). The photoelectrons are collected by an internal movable hemispherical electron energy analyzer that can perform polar
and azimuthal angular scans in UHV. The energy
and momentum resolutions are $\sim$33 meV and $\sim$0.03 \AA$^{-1}$, respectively.



\subsection*{\textit{Ab initio} calculations} \michele{We carried out \textit{ab initio} DFT calculations in a modified 
HSE functional. It improves upon the Perdew, Burke, and Ernzerhof (PBE) \cite{perdew1996generalized,perdew1996phys} exchange-correlation ($xc$) functional 
by the addition of a screened Fock term ($E_{x}^{HF,\textrm{screened}}$), such that the resulting functional reads as
\begin{equation}\label{HSE}
E_{xc}^{HSE}=E_{xc}^{PBE}+\alpha\left(E_{x}^{HF,\textrm{screened}}(\omega)-E_{x}^{PBE,\textrm{screened}}(\omega)\right).
\end{equation}
The screened interaction is written as: $V^\textrm{screened}(r)=\erfc(\omega r)/r$,
where $\erfc$ is the complementary error function,
and $\omega=0.108$ in atomic units, \emph{i.e.} the HSE regular value. In this work, $\alpha$ is instead taken as an adjustable parameter, which depends on the correlation strength of the system.}

We used the \textsc{Quantum Espresso} package \cite{QE-2009,giannozzi2017} 
to perform modified HSE 
calculations for BaNiS$_2$ ($x=1$) and BaCoS$_2$ ($x=0$) \michele{in a plane-waves (PW) basis set}. The geometry of the cell and the internal coordinates
are taken from experiment \cite{Grey1970}. We replaced the core electrons
of the Ni, Co, Ba, and S atoms by norm-conserving
pseudopotentials. For the Ni (Co)
pseudopotential, we used both fully- and scalar-relativistic versions, with 10 (9) valence electrons
and nonlinear core corrections. The Ba pseudopotential
includes the semicore states, while the S pseudopotential has $3s^2
3p^4$ in-valence electrons. We employed a $8 \times 8 \times 8$
electron-momentum grid and a Methfessel-Paxton smearing of 0.01 Ry
for the ${\bf k}$-point integration. The PW cutoff is 60 Ry for the
wave function. The non-local exchange terms of the HSE functional are
computed through the fast implementation of the exact Fock energy
\cite{giannozzi2017}, based on the adaptively compressed exchange
scheme \cite{lin2016}. In the non-local Fock operator evaluation, the
integration over the ${\bf q}$-points is downsampled on a $8\times 8 \times
2$ grid. We applied a half-a-grid shift in the $z$ direction to
minimize the number of nonequivalent momenta in the ${\bf k} +
\textbf{q}$ grid. By
means of the \textsc{Wannier90} code \cite{mostofi2014}, we
performed a Wannier interpolation of the \textit{ab initio} bands
for $x=1$ in
the energy window spanned by the $d-p$ manifold, to accurately
resolve the band structure, chemical potential, and Fermi surface, and
to derive a minimal TB model.

\michele{To successfully deal with the most demanding simulations (HSE functional evaluated in a larger cell with spin resolved orbitals), we supplemented the \textsc{Quantum Espresso} calculations with some performed by means of the \textsc{Crystal17} package\cite{dovesi2018}, particularly suited to efficiently compute the exact exchange operator. In this framework, we used scalar-relativistic Hartree-Fock energy-consistent pseudopotentials by Burkatzki, Filippi, and Dolg\cite{burkatzki2008}, and an adapted VTZ Gaussian basis set, for both Ni and Co. In our \textsc{Crystal17} calculations, the k-grid has been set to a $32 \times 32 \times 32$ dense mesh, with a Fermi smearing of 0.001 Hartree. We cross-checked the \textsc{Crystal17} and \textsc{Quantum Espresso} band structures for the paramagnetic phase of BaNiS$_2$ and BaCoS$_2$, in order to verify the convergence of all relevant parameters in both PW and Gaussian DFT calculations.}
}

\showmatmethods{} 
\acknow{This work was supported by "Investissement d'Avenir" Labex PALM (ANR-10-LABX-0039-PALM), by the Region Ile-de-France (DIM OxyMORE), and by the project CALIPSOplus under Grant Agreement 730872 from the EU Framework Programme for Research and Innovation HORIZON 2020. We acknowledge Beno\^it Baptiste for XRD 
characterization and
Im\`ene Est\`eve for her valuable assistance in the EDS study.
M.C. is grateful to GENCI for the allocation of computer resources under the project N. 0906493. M.F. and A.A. acknowledge support by the European Union, under ERC AdG "FIRSTORM", contract N. 692670.}

\showacknow{} 


\bibliography{library}
\end{document}